\documentclass[aps]{revtex4}
\usepackage[latin1]{inputenc}
\usepackage[francais]{babel}
\usepackage{graphicx}
\usepackage{amsmath}
\usepackage{amsfonts}
\usepackage{amssymb}
\begin{document}
\title{Coherent spin-light-induced mechanisms in the semi-relativistic limit of the self-consistent Dirac-Maxwell equations}
\author{Y. Hinschberger}
\affiliation{Instituto de F\'isica dos materiais da Universidade do Porto, Departamento de F\'isica et astronomica, 687 rua do campo Alegre, 4169-007 Porto, Portugal}
\author{G. Manfredi}
\affiliation{Universit{\'e} de Strasbourg, CNRS, IPCMS UMR 7504, F-67000
Strasbourg, France}
\author{P. -A. Hervieux}
\affiliation{Universit{\'e} de Strasbourg, CNRS, IPCMS UMR 7504, F-67000
Strasbourg, France}
\begin{abstract}
We present a self-consistent mean-field model based on a two-component Pauli-like equation that incorporates quantum and relativistic effects (up to second-order in $1/c$) for both external and internal electromagnetic fields. By taking the semi-relativistic limit of the Dirac-Maxwell equations in the presence of an external electromagnetic field we obtain an analytical expression of a coherent light-induced mean-field Hamiltonian. The latter exhibits several mechanisms that involve the internal mean fields created by all the electrons and the external electromagnetic field (laser). The role played by the light-induced current density and the light-induced second-order charge density acting as sources in Maxwell's equations are clarified. In particular, we identify clearly four different mechanisms involving the spins that may play an important role in coherent ultrafast spin dynamics.
\end{abstract}
\maketitle
\section{Introduction}
Under intense light-matter conditions, relativistic corrections may play a significant role in various systems like dense plasmas \cite{Rid13}, heavy atoms and molecules \cite{John07,Eng03} or condensed-matter systems \cite{Big09}. A theoretical description of the associated charge and spin dynamics can be in principle investigated using relativistic versions of density functional theory (DFT) based on the Dirac-Kohn-Sham equations and relativistic mean-field or Dirac-Hartree-Fock models \cite{Ken07,Raj78,Par84,Rom07}. However, these fully relativistic descriptions involve a four-component Dirac wave function, where the role played by the electronic spinor is concealed, as well as the physical mechanisms that involve the spin degrees of freedom. Such approaches are therefore not useful to validate or uncover new mechanisms in the physics of condensed matter where only electronic 2-component Pauli spinors are considered.

An understanding of ultrafast spin-light interactions is particulary required in the field of femtomagnetism, where the ultrafast demagnetization of ferromagnetic samples induced by femtosecond laser pulses has been studied for almost two decades \cite{Big96,Kir10,Big13} without reaching any consensus regarding the physical mechanisms that underly the quick loss of magnetization. Many relevant proposals have been suggested to explain the ultrafast spin dynamics, ranging from spin-flip scattering involving magnons \cite{Car08}, electrons \cite{Kra09,Gro15} or phonons \cite{Koo10}, to superdiffusive spin current theory \cite{Bat10} or angular momentum transfer with light \cite{Lef09}.

This issue has taken a new turn with an experiment performed on ferromagnetic films \cite{Big09}, showing how to eliminate the ultrafast demagnetization associated to thermal effects and therefore how to have access to the coherent magneto-optical response of the spins. This result indicates that the response of the material induced by a 50-fs laser pulse interacts coherently with the spins to produce a significant magneto-optical effect during the pulse propagation. According to the authors of \cite{Big09}, the observed results may be explained by the relativistic interaction between spins and photons through the Foldy-Wouthuysen (FW) Hamiltonian representing the expansion of the Dirac Hamiltonian at second-order in power of $1/c$ \cite{Str05}. The major role is played by a spin-orbit coupling (SOC) involving the electromagnetic field of the laser pulse which goes beyond the usual SOC due to the electric field of the ions.

To gain a sound understanding of these coherent effects, a theoretical description requires the modeling
of the nonlinear dynamics of a quantum-relativistic system of many interacting electrons excited by an intense and ultrashort
electromagnetic field including all the light-matter terms up to second-order in $1/c$. Unfortunately, an analytical solution
of such many-electron system does not exist, and its numerical solution using \textit{ab-initio} methods is beyond the ability of present-day computers. To circumvent this problem, one may work within a mean-field theory, where the global
effect of the $N$-particle interactions are incorporated in an effective field that acts on a one-particle Hamiltonian. Furthermore, it is worth mentioning that a mean-field derivation including all second-order terms in $1/c$, is a required step before adding further effects such as exchange and correlations.

To achieve this task, in a previous work \cite{DynoI}, we laid the foundations of a two-component self-consistent mean-field model originating from the semi-relativistic limit of the Dirac-Maxwell equations at second-order in $1/c$. It has been shown that a self-consistent theory valid up to second-order in $1/c$ requires the semi-relativistic expansion of the charge and current densities acting as sources in the Maxwell equations, themselves expressed as a power series in the inverse of the speed of light \cite{Manf13}. This model preserves the mathematical structure of the Schr\"odinger or Kohn-Sham equations \cite{Drei90}. In a further work we found that the model is able to describe all the electromagnetic interactions occurring in a two-electron system (as described by the Breit-Pauli Hamiltonian), such as the spin-orbit and spin-other-orbit interactions as well as the spin-spin interaction \cite{DynoII}. Moreover, the model clearly explains how these interactions are created within the single-electron Foldy-Wouthuysen Hamiltonian involving the internal electromagnetic potentials originating from the electronic charge and current densities.

In the present work, we consider the addition of an external electromagnetic field that creates light-induced sources leading to a coherent effective mean-field Hamiltonian. The latter represents the coherent electromagnetic response of the spin-polarized electron gas induced by the external excitation. It contains precisely four identified mechanisms involving the spin degrees of freedom that can be associated to spin-orbit and spin-other-orbit interactions both induced by the external light pulse. A dimensionless analysis based on realistic values of light-matter conditions shows that these coherent effects may play a role in the experimental trends observed in \cite{Big09}. The present results can also enlighten the discussion about the mechanisms underlying the origin(s) of the laser-induced demagnetization.

Finally, let us also mention interesting works that have tried to evaluate the contribution of the relativistic effects proposed in \cite{Big09}. The incorporation of light-induced relativistic terms on hydrogen-like atoms \cite{Von12} or semi-classical Drude-Voigt model \cite{Scud13,Scud15} shows that these effects may produce a magneto-optical contribution, whereas a more recent \textit{ab-initio} investigation on the linear magneto-optical response functions \cite{Opp15} concluded that their contributions are negligible. However, each of the above studies was performed within radically different approximations, so that the debate is still open. We hope that the present work will help focus on the particular spin-light interaction obtained within the Dirac-Maxwell model, which appears to be relevant for addressing the issue of ultrafast coherent magneto-optics.

The paper is organized as follows. In Section II, we present the semi-relativistic Dirac-Maxwell system in the presence of an external electromagnetic field leading to the light-induced mean-field model, including the role of the microscopical sources. We show in Section III that these light-induced effects are not negligible within current light-matter conditions and we perform a detailed analysis of the microscopic mechanisms involving the spins. We conclude in Section IV.
\section{Theory}
\subsection{Semi-relativistic limit of the self-consistent Dirac-Maxwell equations in the presence of an external electromagnetic field}
\begin{figure}[h!]
\begin{center}
\includegraphics[width=8.6cm]{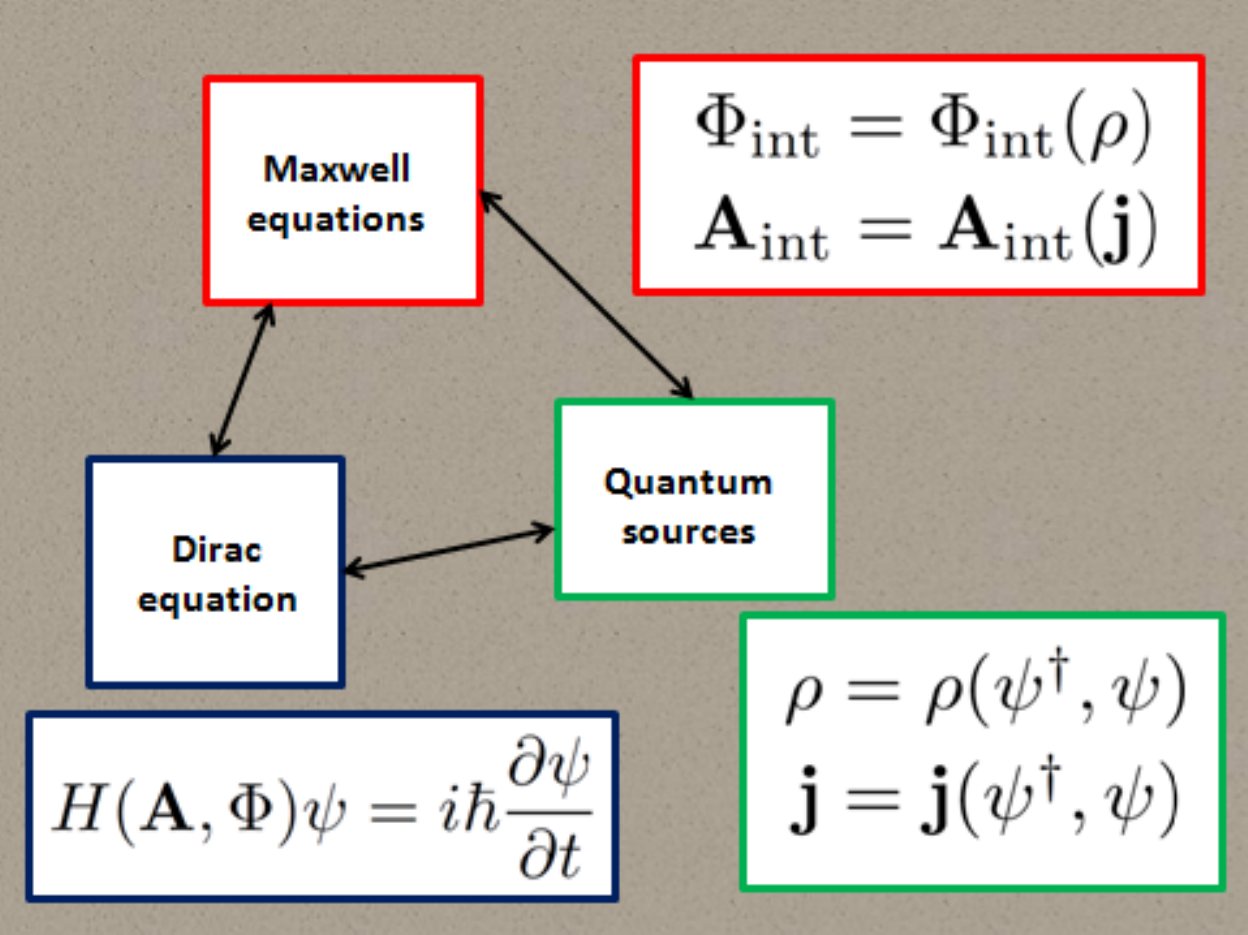}
\end{center}
\caption{(Color online) Scheme of the self-consistent model.}
\label{fig1}
\end{figure}

We consider a many-electron system in the presence of an external electromagnetic field (for instance a laser pulse) where both quantum
and relativistic effects can in principle play a significant role.
In a quantum relativistic mean-field approach, the electron dynamics is governed by the Dirac equation ($q=-e$ with $e>0$),
\begin{eqnarray}
i\hbar\frac{\partial \Psi}{\partial t}&=&\left(c\boldsymbol{\alpha} \cdot  (\mathbf{p}-q\mathbf{A}_{\mathrm{ext}}-q\mathbf{A}_{\mathrm{int}}) \right. \nonumber \\
& & \left. + mc^{2}\beta +q\Phi_{\mathrm{ext}} +q\Phi_{\mathrm{int}}\right)\Psi \;,\label{1}
\end{eqnarray}
where a distinction is made between the \emph{external} potentials $(\Phi_{\mathrm{ext}},\mathbf{A}_{\mathrm{ext}})$ related to the laser pulse and the \emph{internal} potentials $(\Phi_{\mathrm{int}},\mathbf{A}_{\mathrm{int}})$ created by the presence and the motion of all the electrons. The Dirac wave function is a bi-spinor $\Psi=(\phi,\chi)$ where $\phi$ and $\chi$ are, respectively, the electron and positron Pauli spinors and $\boldsymbol{\alpha}$ and $\beta$ are the Dirac's matrix \cite{Str05}. Eq.(\ref{1}) is coupled self-consistently to the Maxwell equations written in terms of the scalar and vector potentials ($\Phi_k$, $\mathbf{A}_k$) ($k$=ext, int) in the
Lorentz gauge ($\boldsymbol{\nabla}\cdot\mathbf{A}_k+\frac{1}{c}\frac{\partial \Phi_k}{\partial t}=0$),
\begin{eqnarray}
\left\{ \begin{array}{c}
-\Delta \Phi_{\mathrm{ext}}+ \frac{1}{c^2}\frac{\partial^2 \Phi_{\mathrm{ext}}}{\partial t^2} = 0  \\
-\Delta \mathbf{A}_{\mathrm{ext}} + \frac{1}{c^2}\frac{\partial^2 \mathbf{A}_{\mathrm{ext}}}{\partial t^2} = 0
\end{array} \right. \;,\; \label{2}
\end{eqnarray}
and
\begin{eqnarray}
\left\{ \begin{array}{c}
-\Delta \Phi_{\mathrm{int}}+ \frac{1}{c^2} \frac{\partial^2 \Phi_{\mathrm{int}}}{\partial t^2} = \frac{q\rho }{\epsilon_{0}} \\
-\Delta \mathbf{A}_{\mathrm{int}} + \frac{1}{c^2} \frac{\partial^2 \mathbf{A}_{\mathrm{int}}}{\partial t^2} = q\mu_0 \mathbf{j}  \end{array}  \;,\;\right. \label{3}
\end{eqnarray}
where the sources are expressed with the 4-component Dirac current density as
\begin{eqnarray}
(\rho c , \mathbf{j} ) &=& c \sum_{i=1}^{N} \left(\Psi_{i}^{\dagger }\Psi_{i}, \Psi_{i}^{\dagger }\boldsymbol{\alpha}\Psi_{i} \right) \label{4} \;.
\end{eqnarray}
Equations (\ref{1}-\ref{4}) constitute a fully relativistic, Lorentz covariant model for describing the quantum dynamics of a system of $N$ interacting electrons in the mean-field approximation. A scheme of this self-consistent model is depicted in Fig. \ref{fig1}.

In the present work, the internal electromagnetic fields are treated in the Coulomb gauge ($\boldsymbol{\nabla}\cdot\mathbf{A}_{\mathrm{int}}=0$) along with the quasi-static approximation $\left(\Delta \mathbf{A}_{\mathrm{int}} \gg  \frac{1}{c^{2}}\frac{\partial^2 \mathbf{A}_{\mathrm{int}}}{\partial t^2}\right)$. In this framework, the Maxwell's equations (\ref{3}) have to be modified and, following the procedure detailed in \cite{Krause_07, Jackson_02} can be expressed in terms of two Poisson-like equations
\begin{eqnarray}
\left\{ \begin{array}{c}
-\Delta \Phi_{\mathrm{int}}= \frac{q\rho }{\epsilon_{0}}  \cr
-\Delta \mathbf{A}_{\mathrm{int}} = q\mu_{0} \mathbf{j}_{\mathrm{T}}  \end{array} \right. \;, \label{5}
\end{eqnarray}
where $\mathbf{j}_{\mathrm{T}}$ is the transverse component of the current density $\mathbf{j}$ (by definition $\mathbf{j}= \mathbf{j}_{\mathrm{T}}
+\mathbf{j}_{\mathrm{L}}$ where $\mathbf{j}_{\mathrm{L}}$ is the longitudinal current density with $\boldsymbol{\nabla} \cdot \mathbf{j}_{\mathrm{T}}=0$ and $\boldsymbol{\nabla} \wedge \mathbf{j}_{\mathrm{L}}=0$ \cite{Jackson_75}). The analytical solutions of Eqs. (\ref{5}) can be expressed as \cite{Krause_07}
\begin{eqnarray}
\Phi_{\mathrm{int}} (\mathbf{x}) &=& \frac{q}{4\pi \epsilon_{0}} \int \frac{d \mathbf{x'} \rho (\mathbf{x'})}{|\mathbf{x}-\mathbf{x'}|} \label{6} \\
\mathbf{A}_{\mathrm{int}}(\mathbf x) &=& \frac{q\mu_{0}}{4\pi}\int d \mathbf x'\left( \frac{\mathbf{j}(\mathbf x')}{2|\mathbf{x}-\mathbf{x'}|}+ \frac{\mathbf{r}(\mathbf{r}\cdot\mathbf{j}(\mathbf x') )}{2|\mathbf{x}-\mathbf{x'}|^{3}} \right) \;, \label{7}
\end{eqnarray}
where $\mathbf{r}\equiv\mathbf{x}-\mathbf{x'}$. In addition, the external fields will be described arbitrary with the quantities $\Phi_{\mathrm{ext}}$ and $\mathbf{A}_{\mathrm{ext}}$.

We consider now the semi-relativistic limit of the Dirac-Maxwell system of Eqs. (\ref{1}), (\ref{2}), and (\ref{5}) valid up to the second-order in powers of $1/c$ where only the electrons should be considered. The Dirac bi-spinor is transformed into a 2-component Pauli spinor $\Psi=(\phi,\chi)\rightarrow \phi$. Using the Foldy-Wouthuysen transformation, the Dirac Hamiltonian in the presence of an electromagnetic field [Eq. (\ref{1})] is expanded to second-order in $1/c$ (also at second-order in $1/m$) \cite{Foldy_59,DynoI,H&H12} to give
\begin{eqnarray}
H &=& mc^2 +\frac{\left(\mathbf{p}-q(\mathbf{A}_{\mathrm{ext}}+\mathbf{A}_{\mathrm{int}} )\right)^{2}}{2m}+q(\Phi_{\mathrm{ext}}+\Phi_{\mathrm{int}})  \nonumber \\
 &-& \frac{q\hbar}{2m}\boldsymbol{\sigma }\cdot \boldsymbol{\nabla}\wedge (\mathbf{A}_{\mathrm{ext}} +\mathbf{A}_{\mathrm{int}}) - \frac{q\hbar^{2}}{8m^{2}c^{2}}\boldsymbol{\nabla }\cdot(\mathbf E_{\mathrm{ext}}+\mathbf{E}_{\mathrm{int}}) \nonumber \\
&-&  \frac{q\hbar}{4m^{2}c^{2}}\boldsymbol{\sigma} \cdot (\mathbf{E}_{\mathrm{ext}}+ \mathbf{E}_{\mathrm{int}})\wedge(\mathbf{p}-q(\mathbf{A}_{\mathrm{ext}}+\mathbf{A}_{\mathrm{int}}))   \label{8}.
\end{eqnarray}
Here, the first term on the right-hand side is the electron rest mass energy, the next two terms are the standard Schr{\"o}dinger Hamiltonian in the presence of an electromagnetic field, the fourth term is the Pauli spin term (Zeeman effect), the $\boldsymbol{\nabla} \cdot \mathbf E$ term is the Darwin term, and the last term represents the spin-orbit coupling (SOC).

In Eq.(\ref{8}) we neglected the term $-\frac{(\mathbf{p}-q\mathbf{A})^{4}}{8m^3c^2}$ which is the first relativistic correction to the electron mass (expansion of the Lorentz factor $\gamma$ to second-order). This assumption is motivated by the fact that this term introduces fourth-order derivatives in the evolution equation, unlike the non-relativistic Schr{\"o}dinger equation which only contains second-order derivatives. Another important point is that this term is of third-order in $1/m$ which is beyond the purpose of this work.

Working within the nonrelativistic Pauli limit also requires to take the low-speed limit of the Maxwell equation. We focus first on Eqs. (\ref{5})  involving the internal potentials produced by the sources. It has already been shown in \cite{DynoI,DynoII} that the charge and current densities can be expanded in powers of $1/c$ with $\rho=\rho^{(0)}+\rho^{(2)}+...$ and $\mathbf{j}=\mathbf{j}^{(0)}+\mathbf{j}^{(2)}+...$ as
\begin{eqnarray}
\rho^{(0)}&=&\phi^{\dagger }\phi \label{9} \;, \\
\mathbf{j}^{(0)}&=&\frac{i\hbar}{2m}\left(\phi \boldsymbol{\nabla}\phi^{\dagger }-\phi^{\dagger }\boldsymbol{\nabla}\phi \right)- \frac{q}{m}\phi^{\dagger }\phi \mathbf{A}  + \frac{\hbar}{2m}\boldsymbol{\nabla}\wedge \left(  \phi^{\dagger } \boldsymbol{\sigma } \phi\right)  \label{10} \;, \\
\rho^{(2)}&=&\frac{\hbar^2}{8m^{2}c^2} \boldsymbol{\nabla}\cdot \boldsymbol{\nabla}(\phi^{\dagger }\phi) -\frac{q\hbar}{4m^2c^2}\boldsymbol{\nabla}\cdot\left((\phi^{\dagger }\boldsymbol{\sigma}\phi)\wedge \mathbf{A}\right) \nonumber \\
& &-\frac{i\hbar^2}{8m^2c^2}\boldsymbol{\nabla}\cdot ( \phi^{\dagger }\boldsymbol{\sigma}\wedge (\boldsymbol{\nabla} \phi)+(\boldsymbol{\nabla}\phi^{\dagger })\wedge\boldsymbol{\sigma}\phi ) ) \label{11} \;, \\
 \mathbf{j}^{(2)}&=& - \frac{q\hbar}{4m^{2}c^{2}}(\phi^{\dagger }\boldsymbol{\sigma}\phi)\wedge \mathbf{E} -\frac{\hbar^2}{8m^{2}c^{2}} \frac{\partial}{\partial t}\boldsymbol{\nabla}(\phi^{\dagger }\phi ) \nonumber \\
& &+  \frac{i\hbar^2}{8m^{2}c^{2}} \frac{\partial}{\partial t} \left(\phi^{\dagger }\boldsymbol{\sigma}\wedge(\boldsymbol{\nabla}\phi)+(\boldsymbol{\nabla}\phi^{\dagger })\wedge\boldsymbol{\sigma}\phi\right) \nonumber \\
& &  +\frac{q\hbar}{4m^{2}c^{2}}\frac{\partial}{\partial t} \left(\phi^{\dagger }\boldsymbol{\sigma}\phi\wedge \mathbf{A} \right). \label{12}
\end{eqnarray}

In order to build a model treating at the same order the equation of motion (Pauli) and the field equations (Maxwell) one should also expand Maxwell's equations (\ref{6}) and (\ref{7}) to the second-order in powers of $1/c$ by writing the electromagnetic potentials as $\Phi_{\mathrm{int}}=\Phi^{(0)}+\Phi^{(2)}+...$ and $\mathbf{A}_{\mathrm{int}}=\mathbf{A}^{(0)}+\mathbf{A}^{(2)}+...$ \cite{Manf13}. Consequently, the Poisson-like equations (\ref{5}) are related to the above sources given in Eqs.(\ref{9}-\ref{12}) as follows
\begin{eqnarray}
\mathbf{A}^{(0)}_{\mathrm{int}}&=&0 \;,\; \label{13}\\
-\Delta \Phi^{(0)}_{\mathrm{int}}&=&\frac{q\rho^{(0)} }{\epsilon_{0}} \;,\;  \label{14} \\
-\Delta \mathbf{A}^{(2)}_{\mathrm{int}} &=&\frac{q\mathbf{j}^{(0)}_{\mathrm{T}} }{\epsilon_{0}c^2} \;,\; \label{15} \\
-\Delta \Phi^{(2)}_{\mathrm{int}}&=& \frac{q \rho^{(2)} }{\epsilon_{0}} \;.\;\label{16}
\end{eqnarray}
The second-order current density $\mathbf{j}^{(2)}$ should be neglected since it would give rise to an internal potential of order $1/c^{4}$. However, that is not the case for the term $\rho^{(2)}$ which is needed to have a complete description. If some external electromagnetic fields are also present (e.g. the laser pulse) these can be assumed to be of zeroth order. The external potentials of Eqs (\ref{2}) can thus be written as $\Phi_{\mathrm{ext}}=\Phi^{(0)}_{\mathrm{ext}}$ and $\mathbf{A}_{\mathrm{ext}}=\mathbf{A}^{(0)}_{\mathrm{ext}}$.

Let us now look at the different terms that make up the current and charge densities. As for the current density $\mathbf{j}^{0}$, the first term on the right-hand side of Eq. (\ref{10}) denotes the orbital charge current, while the last one represents the spin current
\begin{eqnarray}
\mathbf{j}_{\mathrm{orb}}^{(0)}&=&\frac{i\hbar}{2m}\left(\phi \boldsymbol{\nabla}\phi^{\dagger }-\phi^{\dagger }\boldsymbol{\nabla}\phi \right) \label{17} \;,\\
\mathbf{j}_{\mathrm{spin}}^{(0)}&=&\frac{\hbar}{2m}\boldsymbol{\nabla}\wedge \left(  \phi^{\dagger } \boldsymbol{\sigma } \phi\right) \;.\label{18}
\end{eqnarray}
The middle term in Eq.(\ref{10}) reads as $\mathbf{j}_{\mathbf{A}}^{(0)}=- \frac{q}{m}\phi^{\dagger }\phi \mathbf{A}$ and is an electronic current induced by any vector potential (sometimes called the "paramagnetic current"). In the present work, restricted to order $1/c^2$ and $1/m^2$, $\mathbf{j}_{\mathbf{A}}$ can only be induced by the external vector potential, $\mathbf{A}_{\mathrm{ext}}=\mathbf{A}^{(0)}_{\mathrm{ext}}$ being of zeroth order. Indeed, the internal vector potential $\mathbf{A}^{(2)}_{\mathrm{int}}$, being of second-order, would induce a second-order current $\mathbf{j}_{\mathbf{A}^{(2)}_{\mathrm{int}}}^{(0)}\mapsto \mathbf{j}_{\mathbf{A}}^{(2)}$ creating a potential that is of fourth order in $1/c$. Therefore, only the vector potential originating from the external electromagnetic field is considered here, and the associated current is marked with the subscript "field" meaning "field-induced"
\begin{eqnarray}
\mathbf{j}_{\mathrm{field}}^{(0)}&=&- \frac{q}{m}\phi^{\dagger }\phi \mathbf{A}_{\mathrm{ext}} \;.\label{19}
\end{eqnarray}
The above remarks also apply for the second-order density $\rho^{(2)}$, which can be split into the following three types of terms
\begin{eqnarray}
\rho^{(2)}_{\mathrm{orb}}&=&\frac{\hbar^2}{8m^{2}c^2} \boldsymbol{\nabla}\cdot \boldsymbol{\nabla}(\phi^{\dagger }\phi) \;, \label{20} \\
\rho^{(2)}_{\mathrm{spin}}&=&-\frac{i\hbar^2}{8m^2c^2}\boldsymbol{\nabla}\cdot ( \phi^{\dagger }\boldsymbol{\sigma}\wedge (\boldsymbol{\nabla} \phi)+(\boldsymbol{\nabla}\phi^{\dagger })\wedge\boldsymbol{\sigma}\phi)) \;,\label{21} \\
\rho^{(2)}_{\mathrm{field}}&=&-\frac{q\hbar}{4m^2c^2}\boldsymbol{\nabla}\cdot\left((\phi^{\dagger }\boldsymbol{\sigma}\phi)\wedge \mathbf{A}_{\mathrm{ext}} \right) \;.\label{22}
\end{eqnarray}
The quantity  $\rho^{(2)}_{\mathrm{orb}}$ originates from the Darwin term illustrating a correction to the potential energy due to the so-called \textit{Zitterbewegung} (trembling motion of the electron in a volume of size $\lambda_\mathrm{C}^3$ where $\lambda_\mathrm{C}=\frac{h}{mc}$ is the Compton wave-length \cite{Str05}), while $\rho^{(2)}_{\mathrm{spin}}$ and $\rho^{(2)}_{\mathrm{field}}$ are obtained from the spin-orbit interaction. Thereby, it has to be noted that the light-induced term $\rho^{(2)}_{\mathrm{field}}$ involves  also the spin.

As a consequence, the internal fields involving the light-matter operators in the Foldy-Wouthusen Hamiltonian of Eq. (\ref{8}) are split into the following terms
\begin{eqnarray}
\left\{ \begin{array}{ll}
\Phi_{\mathrm{int}} = \Phi^{(0)} +\Phi^{(2)}_{\mathrm{orb}}+\Phi^{(2)}_{\mathrm{spin}}+\Phi^{(2)}_{\mathrm{field}}   \\
\mathbf{A}_{\mathrm{int}} = \mathbf{A}^{(2)}_{\mathrm{orb}}+ \mathbf{A}^{(2)}_{\mathrm{spin}}+\mathbf{A}^{(2)}_{\mathrm{field}} \end{array} \right. \;, \label{23}
\end{eqnarray}
with their analytical expressions using Eqs. (\ref{6}) and (\ref{7}) given by
\begin{eqnarray}
\Phi_{\mathrm{int}}^{(k)} &=& \frac{q}{4\pi \epsilon_{0}}\sum_{i=1}^{N} \int \frac{d \mathbf{x'} \rho ^{(k)}_i(\mathbf{x'})}{|\mathbf{x}-\mathbf{x'}|} \label{24} \\
\mathbf{A}_{\mathrm{int}}^{(l+2)}&=& \frac{q\mu_{0}}{4\pi}\sum_{i=1}^{N} \int d \mathbf x'\left( \frac{\mathbf{j}^{(l)}_i(\mathbf x')}{2|\mathbf{x}-\mathbf{x'}|}+ \frac{\mathbf{r}(\mathbf{r}\cdot\mathbf{j}^{(l)}_i(\mathbf x') )}{2|\mathbf{x}-\mathbf{x'}|^{3}} \right) \label{25}
\end{eqnarray}
where the superscripts $^{(k)}$ and $^{(l)}$ denote respectively the type of sources $\left(^{(k)}=^{(0)},\; ^{(2)}_{\mathrm{orb}},\; ^{(2)}_{\mathrm{spin}},\; ^{(2)}_{\mathrm{field}}\right)$ and $\left(^{(l)}=^{(0)}_{\mathrm{orb}},\;^{(0)}_{\mathrm{spin}},\;^{(0)}_{\mathrm{field}} \right)$, which refer to the equation sets [(\ref{9}),(\ref{20}),(\ref{21}),(\ref{22}))] and [(\ref{17}),(\ref{18}),(\ref{19})]. Finally, by plugging Eq. (\ref{23}) into Eq. (\ref{8}), with the analytical form of Eqs. (\ref{24}) and (\ref{25}), one obtains the low-energy Pauli-equation
\begin{eqnarray}
i\hbar\frac{\partial \phi}{\partial t}&=& \left(mc^2 +\frac{\mathbf{p}^2}{2m} +U^{\mathrm{ext}}+U^{\mathrm{int}}+U^{\mathrm{int}}_{\mathrm{ext}}\right)\phi \;, \label{26}
\end{eqnarray}
which constitutes with Eqs. (\ref{13}), (\ref{14}), (\ref{15}), and (\ref{16}) a self-consistent mean-field model at second-order
in powers of $1/c$ \emph{in the presence of an external electromagnetic field}.

The Pauli Hamiltonian is composed of three groups of terms $U^{\mathrm{ext}}$, $U^{\mathrm{int}}$ and $U^{\mathrm{int}}_{\mathrm{ext}}$. The first ($U_{\mathrm{ext}}$) incorporates the coupling between the electron and the external field. It is just the FW Hamiltonian of a single-electron in the presence of an external electromagnetic field with $\boldsymbol{\nabla } \wedge \mathbf{A}_\mathrm{ext}=\mathbf{B}_\mathrm{ext}$ and $\mathbf{E}_\mathrm{ext} \parallel\mathbf{A}_\mathrm{ext}$:
\begin{eqnarray}
U^{\mathrm{ext}}&=&q\Phi_\mathrm{ext} -\frac{q}{m} \mathbf{A}_\mathrm{ext}\cdot \mathbf{p}+\frac{q^2}{2m}\mathbf{A}_\mathrm{ext}^{2} -\frac{q\hbar}{2m}\boldsymbol{\sigma }\cdot \mathbf{B}_\mathrm{ext} \nonumber \\
&-& \frac{q\hbar^{2}}{8m^{2}c^{2}}\boldsymbol{\nabla }\cdot\mathbf E_\mathrm{ext} - \frac{q\hbar}{4m^{2}c^{2}}\boldsymbol{\sigma} \cdot \mathbf{E}_\mathrm{ext}\wedge\mathbf{p} \;. \label{27}
\end{eqnarray}
Note that even though the external field is treated in the Lorentz gauge, leading to the existence of a term $-\frac{q}{m}\boldsymbol{\nabla }\cdot \mathbf{A}_\mathrm{ext}$, the latter can be neglected in the long wavelength approximation.

The term  $U^{\mathrm{int}}$ is related to the mean internal interactions created by the other electrons of the system, and reads as
\begin{eqnarray}
U^{\mathrm{int}}&=&q\left(\Phi^{(0)} +\Phi^{(2)}_{\mathrm{orb}}+\Phi^{(2)}_{\mathrm{spin}} \right) -\frac{q}{m}\left(\mathbf{A}^{(2)}_{\mathrm{orb}}+ \mathbf{A}^{(2)}_{\mathrm{spin}}\right)\cdot\mathbf{p} \nonumber \\
& & -\frac{q\hbar}{2m}\boldsymbol{\sigma }\cdot \boldsymbol{\nabla}\wedge\left(\mathbf{A}^{(2)}_{\mathrm{orb}}+ \mathbf{A}^{(2)}_{\mathrm{spin}} \right) \nonumber \\
& & + \frac{q\hbar^{2}}{8m^{2}c^{2}}\Delta \Phi^{(0)} + \frac{q\hbar}{4m^{2}c^{2}}\boldsymbol{\sigma} \cdot \boldsymbol{\nabla }\Phi^{(0)} \wedge\mathbf{p} \;.\label{28}
\end{eqnarray}
It has been shown in \cite{DynoII} that the above potential is equivalent to the Breit-Pauli interaction in the Hartree approximation. More importantly, it was shown precisely how the the light-matter operators of the single-electron FW Hamiltonian couple to the different types of internal fields to recover all the electron-electron interactions involved in a two-body system at second-order in $1/c$. Here, we just recall that the first term $q\Phi^{(0)}$ is the usual Hartree term while $q\Phi^{(2)}_{\mathrm{orb}}$ and $\frac{q\hbar^{2}}{8m^{2}c^{2}}\Delta \Phi^{(0)}$ are mean contact terms. The magnetic dipolar term $-\frac{q}{m}\mathbf{A}^{(2)}_{\mathrm{orb}}\cdot\mathbf{p} $ illustrates the coupling between the electron momenta. The term $\frac{q\hbar}{4m^{2}c^{2}}\boldsymbol{\sigma} \cdot \boldsymbol{\nabla }\Phi^{(0)} \wedge\mathbf{p}$ is obviously the spin-orbit interaction with the mean electric field and the second-order potential $q\Phi^{(2)}_{\mathrm{spin}}$ represent the spin-orbit interaction of the mean particle moving around the electron charge. The Zeeman interaction $-\frac{q\hbar}{2m}\boldsymbol{\sigma }\cdot \left(\boldsymbol{\nabla}\wedge\mathbf{A}^{(2)}_{\mathrm{orb}}\right)$ denotes a spin-other-orbit coupling due to the orbital motion, whereas $-\frac{q}{m}\mathbf{A}^{(2)}_{\mathrm{spin}}\cdot\mathbf{p}$ also represents a spin-other-orbit coupling involving the charge motion and the others spins of the system. The last term representing a Zeeman effect related to the spin-current of the system $-\frac{q\hbar}{2m}\boldsymbol{\sigma }\cdot \left(\boldsymbol{\nabla}\wedge\mathbf{A}^{(2)}_{\mathrm{spin}}\right)$ logically gives the spin-spin interaction.

Finally, we focus our attention on the last term of Eq. (\ref{26}) denoted $U^{\mathrm{int}}_{\mathrm{ext}}$. This term represents a light-matter interaction  between the internal mean fields of the system and the external electromagnetic field
\begin{eqnarray}
U^{\mathrm{int}}_{\mathrm{ext}}&=&q\Phi^{(2)}_{\mathrm{field}} -\frac{q}{m}\mathbf{A}^{(2)}_{\mathrm{field}}\cdot \mathbf{p} +\frac{q^2}{m}\mathbf{A}_\mathrm{ext}\cdot\left(\mathbf{A}^{(2)}_{\mathrm{orb}} +\mathbf{A}^{(2)}_{\mathrm{field}}\right)  +\frac{q^2}{m}\mathbf{A}_\mathrm{ext}\cdot \mathbf{A}^{(2)}_{\mathrm{spin}}\nonumber \\
& & -\frac{q\hbar}{2m}\boldsymbol{\sigma }\cdot \left(\boldsymbol{\nabla}\wedge\mathbf{A}^{(2)}_{\mathrm{field}} \right)- \frac{q^2\hbar}{4m^{2}c^{2}}\boldsymbol{\sigma} \cdot \left( \boldsymbol{\nabla }\Phi^{(0)} \wedge \mathbf{A}_\mathrm{ext}\right)  \; . \label{29}
\end{eqnarray}
It represents a coherent light-induced mean field displaying two important properties. First, such term is created by the external field $U^{\mathrm{int}}_{\mathrm{ext}}=U^{\mathrm{int}}_{\mathrm{ext}}(\mathbf{A}_{\mathrm{ext}})$ and is therefore a \emph{coherent interaction}, in the sense that these effects do not exist if the external field is turned off $U^{\mathrm{int}}_{\mathrm{ext}}(\mathbf{A}_{\mathrm{ext}}=0)=0$. Then, all the terms in Eq. (\ref{29}) contain at least one mean internal potential suggesting these effects include all the electrons of the system. It thus represents a macroscopic response of the system to the initial light perturbation. The importance and the outcomes of these terms are discussed in Section III.

Among all the interactions representing $U^{\mathrm{int}}_{\mathrm{ext}}$, it can be seen that three terms act only on the electronic charge: the magnetic dipolar interaction with $-\frac{q}{m}\mathbf{A}^{(2)}_{\mathrm{field}}\cdot \mathbf{p}$ and two other terms originating from the energy-like term $\frac{q^2\mathbf{A}^2}{2m}$ that involves the different vectors $\mathbf{A}$ of the problem: $\frac{q^2}{m}\mathbf{A}_\mathrm{ext}\cdot\mathbf{A}^{(2)}_{\mathrm{orb}}$ and $\frac{q^2}{m}\mathbf{A}_\mathrm{ext}\cdot\mathbf{A}^{(2)}_{\mathrm{field}}$.
Their analytical forms are given below by replacing $\mathbf{A}^{(2)}_{\mathrm{orb}}$ and $\mathbf{A}^{(2)}_{\mathrm{field}}$ with their expressions:
\begin{eqnarray}
-\frac{q}{m}\mathbf{A}^{(2)}_{\mathrm{field}}\cdot \mathbf{p}&=& -\frac{q}{m}\left[ \frac{-q\mu_{0}}{4\pi m}\sum_{i=1}^{N}\int d \mathbf x'\left( \frac{\left(\phi^{\dagger }_i\phi_i q\mathbf{A}_\mathrm{ext} \right)}{2|\mathbf{x}-\mathbf{x'}|}+ \frac{\mathbf{r}\left[\mathbf{r}\cdot\left(\phi_i^{\dagger }\phi_i q\mathbf{A}_\mathrm{ext} \right) \right]}{2|\mathbf{x}-\mathbf{x'}|^{3}} \right) \right]\cdot \mathbf{p} \;, \label{30} \\
\frac{q^2}{m}\mathbf{A}_\mathrm{ext}\cdot\mathbf{A}^{(2)}_{\mathrm{orb}} &=& \frac{q^2}{m}\mathbf{A}_\mathrm{ext}\cdot \frac{q\mu_0}{4\pi 2m} \sum_{i=1}^{N}\int d\mathbf{x} '\left(  \frac{2\phi_i^{\dagger}\mathbf{p}_i\phi_i}{|\mathbf{x}-\mathbf{x'}|}   +\frac{2\mathbf{r}(\phi_i^{\dagger} \mathbf{p}_i\phi_i\cdot\mathbf{r})}{|\mathbf{x}-\mathbf{x'}|^{3}} \right) \;, \label{31} \\
\frac{q^2}{m}\mathbf{A}_\mathrm{ext}\cdot\mathbf{A}^{(2)}_{\mathrm{field}} &=& \frac{q^2}{m}\mathbf{A}_\mathrm{ext}\cdot \frac{-q\mu_{0}}{4\pi m}\sum_{i=1}^{N}\int d \mathbf x'\left( \frac{\left(\phi^{\dagger }_i\phi_i q\mathbf{A}_\mathrm{ext} \right)}{2|\mathbf{x}-\mathbf{x'}|}+ \frac{\mathbf{r}\left[\mathbf{r}\cdot\left(\phi^{\dagger }_i\phi_i q\mathbf{A}_\mathrm{ext} \right) \right]}{2|\mathbf{x}-\mathbf{x'}|^{3}} \right) \;. \label{32}
\end{eqnarray}
The coherent light-induced mean field exhibits also two terms involving directly the spin degrees of freedom. The first one comes from the Zeeman interaction with $-\frac{q\hbar}{2m}\boldsymbol{\sigma }\cdot \left(\boldsymbol{\nabla}\wedge\mathbf{A}^{(2)}_{\mathrm{field}} \right)$ and the second from the spin-orbit operator $- \frac{q^2\hbar}{4m^{2}c^{2}}\boldsymbol{\sigma} \cdot \left( \boldsymbol{\nabla }\Phi^{(0)} \wedge \mathbf{A}_\mathrm{ext}\right)$. They precisely read as
\begin{eqnarray}
-\frac{q\hbar}{2m}\boldsymbol{\sigma }\cdot \left(\boldsymbol{\nabla}\wedge\mathbf{A}^{(2)}_{\mathrm{field}} \right) &=& -\frac{q\hbar}{2m}\boldsymbol{\sigma }\cdot \boldsymbol{\nabla}\wedge \frac{-q\mu_{0}}{4\pi m}\sum_{i=1}^{N}\int d \mathbf x'\left( \frac{\left(\phi_i^{\dagger }\phi_i q\mathbf{A}_\mathrm{ext} \right)}{2|\mathbf{x}-\mathbf{x'}|}+ \frac{\mathbf{r}\left[\mathbf{r}\cdot\left(\phi_i^{\dagger }\phi_i q\mathbf{A}_\mathrm{ext} \right) \right]}{2|\mathbf{x}-\mathbf{x'}|^{3}} \right) \;, \label{33} \\
- \frac{q^2\hbar}{4m^{2}c^{2}}\boldsymbol{\sigma} \cdot \left( \boldsymbol{\nabla }\Phi^{(0)} \wedge \mathbf{A}_\mathrm{ext}\right) &=& -\frac{q^2\hbar }{4m^{2}c^{2}} \frac{q}{4\pi \epsilon_{0}}\boldsymbol{\sigma} \cdot  \boldsymbol{\nabla} \left(\sum_{i=1}^{N} \int \frac{d \mathbf x'\phi_i^{\dagger }\phi_i}{|\mathbf{x}-\mathbf{x'}|} \right) \wedge \mathbf{A}_\mathrm{ext} \;. \label{34}
\end{eqnarray}
Finally, one can also see two terms containing the spins indirectly via the internal fields of the system. The second-order potential energy $q\Phi^{(2)}_{\mathrm{field}}$ has spin-light-induced properties due to the term $\rho^{(2)}_{\mathrm{field}}$ [see Eq. (\ref{22})] and the operator $\frac{q^2}{m}\mathbf{A}_\mathrm{ext}\cdot \mathbf{A}^{(2)}_{\mathrm{spin}}$ couples the external vector potential with the one of the spin system. In both cases, the spin represented by the Pauli matrix $\boldsymbol{\sigma}$ remains inside the integrals
\begin{eqnarray}
q\Phi^{(2)}_{\mathrm{field}} &=& q\frac{q}{4\pi \epsilon_{0}} \left(-\frac{\hbar}{4m^2c^2} \right) \sum_{i=1}^{N}\int \frac{d \mathbf{x'} \boldsymbol{\nabla}\cdot\left((\phi_i^{\dagger }\boldsymbol{\sigma}\phi_i)\wedge q\mathbf{A}_{\mathrm{ext}}\right)}{|\mathbf{x}-\mathbf{x'}|} \;, \label{35} \\
\frac{q^2}{m}\mathbf{A}_\mathrm{ext}\cdot \mathbf{A}^{(2)}_{\mathrm{spin}}&=& \frac{q^2}{m}\mathbf{A}_\mathrm{ext}\cdot \frac{q\mu_{0}}{4\pi}\sum_{i=1}^{N} \int d \mathbf x'\left( \frac{\boldsymbol{\nabla}\wedge \left(  \phi_i^{\dagger } \boldsymbol{\sigma } \phi_i\right)}{2|\mathbf{x}-\mathbf{x'}|}+ \frac{\mathbf{r}\left[\mathbf{r}\cdot\boldsymbol{\nabla}\wedge \left(  \phi_i^{\dagger } \boldsymbol{\sigma } \phi_i\right)\right]}{2|\mathbf{x}-\mathbf{x'}|^{3}} \right)  \;. \label{36}
\end{eqnarray}
A detailed analysis of the last four terms is performed in Section III.C. We show quickly in the next paragraph, how the latter interactions are related to those of the Breit-Pauli Hamiltonian.
\subsection{Equivalence with the Breit-Pauli Hamiltonian within the Hartree mean field approximation}
The Breit-Pauli Hamiltonian describes the interaction between two moving electrons at second-order in $1/c$. It has its origin in the non-relativistic limit of the Breit Hamiltonian $H^{B}$ describing the retardation effects on the electromagnetic energy between two electrons in the Dirac's formalism: $H^{B}=\bar{e}^{2}\frac{\boldsymbol{\alpha}_{i}\cdot \boldsymbol{\alpha}_{j}}{r_{ij}} + \bar{e}^{2}\frac{ (\boldsymbol{\alpha}_{i}\cdot \mathbf{r}_{ij})(\boldsymbol{\alpha}_{j}\cdot \mathbf{r}_{ij})}{r_{ij}^{3}}$ \cite{Bre29,Bet77,LanIV} where $\bar{e}^2=\frac{q^2}{4\pi\epsilon_0}$. The latter can be built from the classical Darwin Lagrangian $L^D= -\bar{e}^{2}\frac{\mathbf{v}_{i} \cdot \mathbf{v}_{j}}{r_{ij}} - \bar{e}^{2}\frac{ (\mathbf{v}_{i}\cdot \mathbf{r}_{ij})(\mathbf{v}_{j}\cdot \mathbf{r}_{ij})}{r_{ij}^{3}}$ \cite{Dar20}, which represents the classical energy of two moving charges $U=\frac{1}{2}\sum_{i \neq j} \left(q_i\Phi_j -q_i\mathbf{v}_i\cdot \mathbf{A}_j \right)$ where $\Phi$ and $\mathbf A$ are obtained by expanding to second-order in $1/c$ the Lienard-Wieckert potentials \cite{LanII}. The Breit-Pauli Hamiltonian $H_{ij}^{BP}$ completes the classical description by adding the quantum and relativistic properties due to the electron spins
\begin{widetext}
\begin{eqnarray}
H_{ij}^{BP}&=&-\frac{\pi\hbar^{2}\bar{e}^{2}}{m^{2}c^{2}}\delta (r_{ij})-\frac{\bar{e}^{2}}{2m^{2}c^{2}}\left( \frac{\mathbf{p}_{i}\cdot \mathbf{p}_{j}}{r_{ij}} + \frac{\mathbf{r}_{ij}\cdot (\mathbf{p}_{j}\cdot \mathbf{r}_{ij})\mathbf{p}_{i}}{r_{ij}^{3}}\right) \nonumber \\
& &+\frac{\hbar \bar{e}^{2}}{4m^{2}c^{2}} \left( \boldsymbol{\sigma}_{j}\cdot \frac{\mathbf{r}_{ij}}{r_{ij}^3}\wedge (\mathbf{p}_j-2\mathbf{p}_i)  -\boldsymbol{\sigma}_{i}\cdot \frac{\mathbf{r}_{ij}}{r_{ij}^3}\wedge (\mathbf{p}_i-2\mathbf{p}_j) \right) \nonumber \\
& & -\frac{\hbar \bar{e}^{2}}{4m^{2}c^{2}}\left( -8\pi\frac{ \boldsymbol{\sigma}_{i}\cdot\boldsymbol{\sigma}_{j}}{3}\delta (r_{ij})
-\frac{ \boldsymbol{\sigma}_{i}\cdot\boldsymbol{\sigma}_{j}}{r_{ij}^{3}} +3 \frac{ (\boldsymbol{\sigma}_{i}\cdot\mathbf{r}_{ij})(\boldsymbol{\sigma}_{j}\cdot \mathbf{r}_{ij})}{r_{ij}^{5}}\right)\;,  \label{37}
\end{eqnarray}
where the first line of Eq. (\ref{37}) denotes "spin-free" terms through a contact operator and the coupling between the electronic momenta, the second line represents the "spin-orbit" and "spin-other-orbit" interactions, and the last line illustrates the "spin-spin" interaction. It has been shown in \cite{DynoII}, that for a system of $N$ interacting electrons, the Breit-Pauli operators in the mean-field Hartree approximation are equivalent to the operators of $U^{\mathrm{int}}$ given by Eq. (\ref{28}).
When one considers an additional external field ($\Phi_\mathrm{ext},\mathbf{A}_\mathrm{ext}$), the Breit-Pauli Hamiltonian is modified into $H_{ij}^{BP}\mapsto H_{ij}^{BP}(\mathbf{A}_\mathrm{ext})$. Indeed, by performing a Foldy-Whouthuysen transformation on a Dirac-Breit two-electron system, one can show that the modification brought by the external field up to second-order in $1/c$ only requires the canonical substitution $\mathbf{p}_i \mapsto \mathbf{p}_i-q_i\mathbf{A}_\mathrm{ext}$ and $\mathbf{p}_j \mapsto \mathbf{p}_j-q_j\mathbf{A}_\mathrm{ext}$ \cite{H&H15}. Consequently, these modifications only affect the "spin-free", "spin-orbit" and "spin-other-orbit" terms of Eq. (\ref{37}) leading to
\begin{eqnarray}
H_{ij}^{BP}(\mathbf{A}_\mathrm{ext})&=&H_{ij}^{BP}+\frac{\bar{e}^{2}}{2m^{2}c^{2}}\left( \frac{\mathbf{p}_{i}\cdot q_j\mathbf{A}_\mathrm{ext}}{r_{ij}} + \frac{\mathbf{r}_{ij}\cdot (q_j\mathbf{A}_\mathrm{ext}\cdot \mathbf{r}_{ij})\mathbf{p}_{i}}{r_{ij}^{3}}  \right) \nonumber \\
& &  +\frac{\bar{e}^{2}}{2m^{2}c^{2}}\left(\frac{q_i\mathbf{A}_\mathrm{ext}\cdot \mathbf{p}_{j}}{r_{ij}} + \frac{\mathbf{r}_{ij}\cdot (\mathbf{p}_{j}\cdot \mathbf{r}_{ij})q_j\mathbf{A}_\mathrm{ext}}{r_{ij}^{3}}   \nonumber \right) \\
& & +\frac{\bar{e}^{2}}{2m^{2}c^{2}}\left(-\frac{q_iq_j\mathbf{A}_\mathrm{ext}^2}{r_{ij}}  -q_iq_j\frac{\mathbf{r}_{ij}\cdot (\mathbf{A}_\mathrm{ext}\cdot \mathbf{r}_{ij})\mathbf{A}_\mathrm{ext}}{r_{ij}^{3}}  \right) \nonumber \\
& &+\frac{\hbar \bar{e}^{2}}{4m^{2}c^{2}} \left( \boldsymbol{\sigma}_{j}\cdot \frac{\mathbf{r}_{ij}}{r_{ij}^3}\wedge (2q_i\mathbf{A}_\mathrm{ext}-q_j\mathbf{A}_\mathrm{ext})  -\boldsymbol{\sigma}_{i}\cdot \frac{\mathbf{r}_{ij}}{r_{ij}^3}\wedge (2q_j\mathbf{A}_\mathrm{ext}-q_i\mathbf{A}_\mathrm{ext}) \right) \;.\label{38}
\end{eqnarray}
\end{widetext}
In this case, the exact Hamiltonian of the $N$ interacting electrons in the presence of an external electromagnetic field at second-order in powers of $1/c$ is given by
\begin{eqnarray}
H=\sum_{i=1}^{N}\frac{\mathbf{p}^2_{i}}{2m}+mc^2+U^{\mathrm{ext}}_{\sigma_i,p_i} +\frac{1}{2}\sum_{ j \neq i }\frac{\bar{e}^2}{r_{ij}}+H_{ij}^{BP}(\mathbf{A}_{\mathrm{ext}}) \;. \nonumber
\end{eqnarray}
The modifications that bring the external field are twofold: i) an action on each electron given by $U_{\sigma_i,p_i}^{ext}$ which is just the FW Hamiltonian at second-order in $1/m$; ii) the addition of the extra-terms given by Eq. (\ref{38}). By taking the total wave-function of the system in the Hartree approximation $\phi(\mathbf{r}_{1},...,\mathbf{r}_{N})=\phi_{1}(\mathbf{r}_{1})\phi_{2}(\mathbf{r}_{2})...\phi_{N}(\mathbf{r}_{N})$ and using the Lagrange method of undetermined multipliers, one obtains the Hartree-Breit-Pauli equations for a spinor $\phi_i(\mathbf{r}_{i})\equiv\phi$ which is a solution of the single-particle Pauli equation: $ \left(\frac{\mathbf{p}^2}{2m}+ mc^2+ U^{\mathrm{ext}}+ U^{BP}_{\mathrm{eff}}+ U^{BP}_{\mathrm{eff(A)}}\right)\phi=i\hbar \frac{\partial \phi}{\partial t}$, where $U^{BP}_{\mathrm{eff}}$ is equivalent to Eq. (\ref{28}) and $U^{BP}_{\mathrm{eff(A)}}$ represents the contribution of the new terms of Eq. (\ref{38}). The latter are split into "spin-free" terms $(sf)$ and "spin" terms ($\sigma$) as $U^{BP}_{\mathrm{eff(A)}}=^{sf}U^{BP}_{\mathrm{eff(A)}}+ ^{\sigma}U^{BP}_{\mathrm{eff(A)}}$. The expressions of the mean fields given by the first three spin-free terms of Eq. (\ref{38}) read respectively as
\begin{eqnarray}
^{sf}U^{BP}_{\mathrm{eff(A)}}&=& \underbrace{\frac{\bar{e}^{2}}{2m^{2}c^{2}}\left( \sum_{i\neq j}\int d \mathbf{x}' \phi_j^{\dagger}(\mathbf{x}') \left( \frac{q_j\mathbf{A}_\mathrm{ext}}{| \mathbf{x}-\mathbf{x}' |} + \frac{\mathbf{r}(q_j\mathbf{A}_\mathrm{ext}\cdot \mathbf{r})}{| \mathbf{x}-\mathbf{x}' |^{3}} \right)\phi_j (\mathbf{x}')\right) \cdot \mathbf{p} }_{-\frac{q}{m}\mathbf{A}^{(2)}_{\mathrm{field}}\cdot \mathbf{p}\text{: Eq.} (\ref{30})} \;,\nonumber \\
& & \underbrace{+\frac{q\bar{e}^{2}}{2m^{2}c^{2}} \mathbf{A}_\mathrm{ext} \cdot  \sum_{i\neq j}\int d \mathbf{x}' \phi_j^{\dagger} (\mathbf{x}')\left( \frac{\mathbf{p}_{j}}{| \mathbf{x}-\mathbf{x}' |} + \frac{\mathbf{r}(\mathbf{p}_{j}\cdot \mathbf{r})}{| \mathbf{x}-\mathbf{x}' |^{3}}  \right)\phi_j (\mathbf{x}')}_{\frac{q^2}{m}\mathbf{A}_\mathrm{ext}\cdot\mathbf{A}^{(2)}_{\mathrm{orb}}\text{: Eq.} (\ref{31})} \nonumber \;, \\
& & \underbrace{+\frac{q\bar{e}^{2}}{2m^{2}c^{2}} \mathbf{A}_\mathrm{ext} \cdot \sum_{i\neq j}\int d \mathbf{x}' \phi_j^{\dagger} (\mathbf{x}')\left( -\frac{q_j\mathbf{A}_\mathrm{ext}}{| \mathbf{x}-\mathbf{x}' |}    -\frac{\mathbf{r}(q_j\mathbf{A}_\mathrm{ext}\cdot \mathbf{r})}{| \mathbf{x}-\mathbf{x}' |^3} \right)\phi_j (\mathbf{x}')}_{\frac{q^2}{m}\mathbf{A}_\mathrm{ext}\cdot\mathbf{A}^{(2)}_{\mathrm{field}}\text{: Eq.} (\ref{32})} \;,\label{39}
\end{eqnarray}
where the underbraces indicate the correspondence of each term with the "spin-free" terms obtained in Eqs. (\ref{30}), (\ref{31}) and (\ref{32}). The case of the "spin" terms is a bit more tricky but does not present major difficulties. These mean-field interactions are shown below in Eq. (\ref{40}) and correspond to the four spin terms  of Eq. (\ref{38}) respectively taken from the left to the right of the last line
\begin{eqnarray}
^{\sigma}U^{BP}_{\mathrm{eff(A)}}&=&  \underbrace{\frac{ q\hbar \bar{e}^{2}}{2m^{2}c^{2}} \mathbf{A}_\mathrm{ext} \cdot \sum_{i\neq j}\int d \mathbf{x}' \phi_j^{\dagger}(\mathbf{x}') \left( \boldsymbol{\sigma}_{j}\wedge \frac{\mathbf{r}}{\mathbf{r}^3}\right)\phi_j (\mathbf{x}')}_{\text{Spin-other-orbit:} \frac{q^2}{m}\mathbf{A}_\mathrm{ext}\cdot \mathbf{A}^{(2)}_{\mathrm{spin}}\text{: Eq.} (\ref{36})} -\underbrace{\frac{ \hbar \bar{e}^{2}}{4m^{2}c^{2}} \sum_{i\neq j}\int d \mathbf{x}' \phi_j^{\dagger} (\mathbf{x}') \left( \boldsymbol{\sigma}_{j}\cdot \frac{\mathbf{r}}{\mathbf{r}^3}\wedge q_j \mathbf{A}_\mathrm{ext})\right)\phi_j (\mathbf{x}')}_{\text{Spin-orbit:} q\Phi^{(2)}_{\mathrm{field}}\text{: Eq.} (\ref{35})} \nonumber \\
& &  -\underbrace{\frac{\hbar \bar{e}^{2}}{2m^{2}c^{2}} \boldsymbol{\sigma} \cdot \sum_{i\neq j}\int d \mathbf{x}' \phi_j^{\dagger}(\mathbf{x}') \left( \frac{\mathbf{r}}{\mathbf{r}^3}\wedge q_j\mathbf{A}_\mathrm{ext}\right) \phi_j(\mathbf{x}')}_{\text{Spin-other-orbit:}-\frac{q\hbar}{2m}\boldsymbol{\sigma }\cdot \left(\boldsymbol{\nabla}\wedge\mathbf{A}^{(2)}_{\mathrm{field}} \right)\text{: Eq.} (\ref{33})}   + \underbrace{\frac{q\hbar \bar{e}^{2}}{4m^{2}c^{2}} \boldsymbol{\sigma}\cdot \left(\sum_{i\neq j}\int d \mathbf{x}' \phi_j^{\dagger}(\mathbf{x}')  \frac{\mathbf{r}}{\mathbf{r}^3} \phi_j(\mathbf{x}') \right)\wedge \mathbf{A}_\mathrm{ext}}_{\text{Spin-orbit:}- \frac{q^2\hbar}{4m^{2}c^{2}}\boldsymbol{\sigma} \cdot ( \boldsymbol{\nabla }\Phi^{(0)} \wedge \mathbf{A}_\mathrm{ext})\text{: Eq.} (\ref{34})} \nonumber \,.\;\\ \label{40}
\end{eqnarray}
The equivalence with the spin terms of Eq. (\ref{29}) is indicated in the underbraces and can be understood as follows. In Eq. (\ref{38}), the four spin operators refer to particle $i$ or $j$ with the subscripts indicated in the spins ($\boldsymbol{\sigma}_{i},\boldsymbol{\sigma}_{j}$) and the electronic charges ($q_i,q_j$). When applying the Lagrange method of undetermined multipliers $\delta \left<\phi(\mathbf{r}_{1},..,\mathbf{r}_{N})|U^{BP}_{\mathrm{eff(A)}}|\phi(\mathbf{r}_{1},..,\mathbf{r}_{N})\right>=0$, the operators related to particle $i$ ($\boldsymbol{\sigma}_{i}, q_i \mathbf{A}_{\mathrm{ext}}$) are taken out of the integral becoming ($\boldsymbol{\sigma}, q\mathbf{A}_{\mathrm{ext}}$), while those of particle $j$ ($\boldsymbol{\sigma}_{j}, q_j\mathbf{A}_{\mathrm{ext}}$) remain inside the integral. Hence the first spin-other-orbit term involving $(\boldsymbol{\sigma}_{j}, q_i\mathbf{A}_{\mathrm{ext}})$ in Eq. (\ref{38}) leads to a term looking like $\approx q\mathbf{A}_{\mathrm{ext}}\int d\mathbf{x}'...\boldsymbol{\sigma}_{j}$ which can obviously be identified to $\frac{q^2}{m}\mathbf{A}_{\mathrm{ext}}\cdot \mathbf{A}^{(2)}_{\mathrm{spin}}$ of Eq. (\ref{36}). Then, one can see that for the second term, exhibiting a spin-orbit interaction, both quantities $\boldsymbol{\sigma}_{j}$ and  $q_j\mathbf{A}_{\mathrm{ext}}$ will be kept inside the integral and can only correspond to the term $q\Phi^{(2)}_{\mathrm{field}}$ of Eq. (\ref{35}). The same procedure is used to attribute the spin-other-orbit term $(\boldsymbol{\sigma}_{i}, q_j\mathbf{A}_{\mathrm{ext}})$ to $-\frac{q\hbar}{2m}\boldsymbol{\sigma }\cdot \left(\boldsymbol{\nabla}\wedge\mathbf{A}^{(2)}_{\mathrm{field}} \right)$ and the last spin-orbit term $(\boldsymbol{\sigma}_{i}, q_i\mathbf{A}_{\mathrm{ext}})$ to $- \frac{q^2\hbar}{4m^{2}c^{2}}\boldsymbol{\sigma} \cdot ( \boldsymbol{\nabla }\Phi^{(0)} \wedge \mathbf{A}_\mathrm{ext})$. \\

The equivalence of the light-induced mean field originating from the Dirac-Maxwell equations with the Breit-Pauli Hamiltonian in the presence of an external electromagnetic field is still valid. This fact may appear less surprising if one notes that both approaches are performed under the same conditions, namely the Coulomb gauge and the quasi-static approximation. Anyway, following this particular result, one can claim that the light-matter operators involving the spin can be attributed to the spin-orbit and spin-other-orbit interactions, both induced by the external electromagnetic field.
\section{Detailed analysis}
\subsection{Order of magnitude of the light-induced mean field terms}
The Pauli-like Hamiltonian in Eq. (\ref{26}) exhibits three types of terms namely the interaction with the external field $U^{\mathrm{ext}}$, the  mean internal interactions $U^{\mathrm{int}}$ and the semi-relativistic light-induced mean field $U^{\mathrm{int}}_{\mathrm{ext}}$. We propose, as a first approximation, to express these Hamiltonians with dimensionless parameters involving the internal and external parameters of the system. The first term $U^{\mathrm{ext}}$ is the FW Hamiltonian at second-order in $1/c$. In their seminal paper, Foldy and Wouthuysen explained that the dimensionless quantities in their transformation were $\frac{h}{mc}\boldsymbol{\nabla}$ and $\frac{h}{mc^2} \frac{\partial }{\partial t}$ \cite{Foldy_59}. By considering a time-dependent external electromagnetic fields, the spatial (time) derivative can be replaced by $\lambda^{-1}$ ($\omega$) and the dimensionless quantities read as $\frac{\lambda_\mathrm{C}}{\lambda}$  ($\frac{\omega_\mathrm{C}}{\omega}$) where $\lambda_\mathrm{C}=\frac{h}{mc}$ is the Compton wavelength and $\omega_\mathrm{C} =\frac{2\pi c}{\lambda_\mathrm{C}}$ is the Compton frequency. Then, by using the following relations involving the external electromagnetic field quantities $|B_{\mathrm{ext}}|\approx|A_{\mathrm{ext}}/\lambda|$, $|E_{\mathrm{ext}}|\approx |\Phi_{\mathrm{ext}}/\lambda|$, $|\Phi_{\mathrm{ext}}|\approx |cA_{\mathrm{ext}}|$ and $|E_{\mathrm{ext}}|=|cB_{\mathrm{ext}}|$, one may express $U^{\mathrm{ext}}$ as
\begin{eqnarray}
U^{\mathrm{ext}}&\approx & e\Phi_{\mathrm{ext}}\left(1+\frac{\lambda_\mathrm{C}}{\lambda}+\left(\frac{\lambda_\mathrm{C}}{\lambda}\right)^2 +\mathcal{O}\left( c^{-3}\right)\right) \;,\; \label{41}
\end{eqnarray}
where $\Phi_{\mathrm{ext}}$ is the scalar potential of the external electromagnetic field. In the case of the internal mean field $U^{\mathrm{int}}$, which also represents an expansion to second-order in $1/c$, the electric field is due to the Coulomb interaction and the wavelength of the external field has to be replaced by a characteristic electronic distance. By taking the distance between two interacting electrons $r_{ij}$, one can see that the operators of $U^{\mathrm{int}}$ can be expressed as
\begin{eqnarray}
U^{\mathrm{int}} &\approx & \frac{N\bar{e}^2}{r_{ij}}\left(1+ \left(\frac{\lambda_\mathrm{C}}{r_{ij}}\right)^2 +\mathcal{O}\left( c^{-3}\right)\right) \;,\; \label{42}
\end{eqnarray}
where $N$ is the number of electrons in the system. Finally, the light-induced mean-field $U^{\mathrm{int}}_{\mathrm{ext}}$, which belongs exclusively to the second-order in $1/c$, is composed of both internal and external properties and thus reads
\begin{eqnarray}
U^{\mathrm{int}}_{\mathrm{ext}} &\approx &  \frac{N\bar{e}^2}{r_{ij}} \left(\frac{\lambda_\mathrm{C}}{r_{ij}}\right)  \left(\frac{e\Phi_{\mathrm{ext}}}{mc^2}\right) +\mathcal{O}\left( c^{-3}\right) \;. \label{43}
\end{eqnarray}
We want to estimate the importance of the light-induced mean field $U^{\mathrm{int}}_{\mathrm{ext}} $. Being itself of the second-order in $1/c$ it has to be compared to the second-order energy corrections of the internal mean field $U^{\mathrm{int}}$. Indeed, it is the spin-orbit, the spin-other-orbit and the spin-spin interactions that contribute to the magnetic ordering of the $N$-electron system. The latter are represented by the second term in Eq. (\ref{42}) and can be written as $U^{\mathrm{int}(2)}_{\mathrm{ext}}\approx \frac{N\bar{e}^2}{r_{ij}} \left(\frac{\lambda_\mathrm{C}}{r_{ij}}\right)^2$. The light-induced mean field can modify the internal ordering without necessary reaching the ionization regime for which one needs to have at least the energy of the Coulomb interaction $U^{\mathrm{int}(0)}\approx \frac{N\bar{e}^2}{r_{ij}}$. Consequently, using $\Phi_{\mathrm{ext}}=(E_{\mathrm{ext}}\lambda)$ we define a yield parameter $\eta$ that reads
\begin{eqnarray}
\eta =\frac{U^{\mathrm{int}}_{\mathrm{ext}}}{U^{\mathrm{int}(2)}}= \frac{r_{ij}}{\lambda_\mathrm{C}}\frac{e E_{\mathrm{ext}} \lambda }{mc^2}  \;. \label{44}
\end{eqnarray}
With a basic dimensionless approach, the yield parameter finally depends only on the intrinsic electron wavelength (the Compton wavelength: $\lambda_\mathrm{C}=2.42 \times 10^{-12}$ m), a characteristic electronic distance ($r_{ij}$) and the amplitude and wavelength of the external electromagnetic field.

To get a quantitative estimation, we propose to focus on the experiment where the coherent ultrafast magneto-optical measurements were performed on ferromagnetic Nickel thin film \cite{Big09}. As mentioned in the introduction, the authors suggested that the observed trends originate from a relativistic coupling between spin and photons through a Zeeman effect and a spin-orbit coupling involving the electromagnetic fields of the laser as well as the coherent magnetic response of all the interacting spin-electron gas (including all the field terms up to second-order in $1/c$). The present model accurately describes these effects at least in the mean-field approximation (without exchange and correlation effects). In reference \cite{Big09}, the 50-fs laser field was centered at $\lambda \approx 800$ nm with an intensity $\mathcal{E}_0$ around $\mathcal{E}_0=1$ mJ/cm$^{2}$ and the thickness of the Nickel film was $7.5$ nm. Using the relation $\frac{c\epsilon_0 E^2_{\mathrm{ext}}}{2}=\frac{10 \times \mathcal{E}_0(\text{mJ/cm}^{2})}{\Delta t}$ \cite{Scud13} one can estimate $E_{\mathrm{ext}}\approx 4\times 10^{8}$ V/m. The choice of an electronic distance is maybe more difficult. One could take  $r_{ij}\approx 10^{-10}$ m, or even larger since $r_{ij}$ could also represent the typical inter-electronic distance in the Nickel film. For instance with $r_{ij}$ included in the interval $r_{ij} \in [10^{-10} \text{m}, 3\times 10^{-10} \text{m}]$ one obtains $\eta \in [3\%,9\%]$. Even using the lower limit, the result is not so negligible considering the rough approximation provided by the dimensionless analysis. Hence, it would be relevant to perform a sophisticated and rigorous analysis based on a numerical study, which should be able to give a more precise estimation of $\eta$.

However, these effects may play a really important role by employing a larger field amplitude. With an electric field amplitude of $E_{\mathrm{ext}}= 10^{10}$ V/m, a value that is currently available in laboratories, the value of $\eta$ can easily reach $\eta \approx 65 \% $ even with $r_{ij}\approx 10^{-10}$ m. Therefore, these coherent effects appear to be somehow important, and we thus perform a detailed analysis of the related mechanisms in the following section.
\subsection{Mechanisms involving the coherent spin dynamics}
As explained in the introduction, the origin of the quick loss of magnetization following the interaction of a ferromagnetic sample with an ultrafast femtosecond laser pulse is still under active debate. The demagnetization process occurs within two kinds of physical interactions. The first is related to the electromagnetic ordering of the system induced by the polarization of the external electromagnetic field, while the second is linked to its internal disorder created by the associated thermal effects. In the last case, the heat filled by the laser generates an increase of the system temperature, and the thermal agitation modifies randomly each spin-orientation leading to a diminution on the average magnetization of the sample.

The coherent magneto-optical signal extracted from the experiment performed in \cite{Big09} shows that light-induced coherent effects play an important role in the first few femtoseconds of the demagnetization process. It is legitimate to ask what are the main physical mechanisms underlying these effects. For that purpose, let us analyze the spin terms of the Pauli-like Hamiltonian of Eq. (\ref{26}) that only involve the external electromagnetic field and the spin degrees of freedom $^{\sigma}U$. One may distinguish
\begin{eqnarray}
^{\sigma}U&=&^{\sigma}U^{\mathrm{ext}}+^{\sigma}U^{\mathrm{int}}_{\mathrm{ext}} \;,\;  \nonumber
\end{eqnarray}
where $^{\sigma}U^{\mathrm{ext}}$ represents the \emph{direct} coupling between the spin and the laser through the Zeeman interaction and a laser-induced spin-orbit coupling with
\begin{eqnarray}
^{\sigma}U^{\mathrm{ext}}&=& -\frac{q\hbar}{2m}\boldsymbol{\sigma }\cdot \mathbf{B}_\mathrm{ext} - \frac{q\hbar}{4m^{2}c^{2}}\boldsymbol{\sigma} \cdot \mathbf{E}_\mathrm{ext} \wedge\mathbf{p} \;,\; \label{45}
\end{eqnarray}
and $^{\sigma}U^{\mathrm{int}}_{\mathrm{ext}}$ stands for the \emph{indirect} coupling between the spin and the laser which reads as
\begin{eqnarray}
^{\sigma}U^{\mathrm{int}}_{\mathrm{ext}}&=&-\frac{q\hbar}{2m}\boldsymbol{\sigma }\cdot \left(\boldsymbol{\nabla}\wedge\mathbf{A}^{(2)}_{\mathrm{field}} \right)- \frac{q^2\hbar}{4m^{2}c^{2}}\boldsymbol{\sigma} \cdot \left( \boldsymbol{\nabla }\Phi^{(0)} \wedge \mathbf{A}_\mathrm{ext}\right)\nonumber \\
& & q\Phi^{(2)}_{\mathrm{field}}+\frac{q^2}{m}\mathbf{A}_\mathrm{ext}\cdot \mathbf{A}^{(2)}_{\mathrm{spin}}  \;. \label{46}
\end{eqnarray}
The operators in Eq. (\ref{45}) illustrate a direct interaction between the laser electromagnetic field and the electron spin. The latter have already been mentioned in other works \cite{Big09,Von12,H&H12} and do not constitute the purpose of the present work.

Let us now focus on the elements of Eq. (\ref{46}), which can be separated into two types of interactions:

- (A) The first two terms exhibit explicitly the Pauli spinor $\boldsymbol{\sigma }$ and involve a Zeeman-like interaction $-\frac{q\hbar}{2m}\boldsymbol{\sigma}\cdot \left(\boldsymbol{\nabla}\wedge\mathbf{A}^{(2)}_{\mathrm{field}} \right) $ and a SOC-like operator $- \frac{q^2\hbar}{4m^{2}c^{2}}\boldsymbol{\sigma} \cdot ( \boldsymbol{\nabla }\Phi^{(0)} \wedge \mathbf{A}_\mathrm{ext})$. They represent the mean-charge responses (characterized by $(\Phi^{(0)},\mathbf{A}^{(2)}_{\mathrm{field}})$) acting on the electron spin.

- (B) The two others $q\Phi^{(2)}_{\mathrm{field}}$ and $\frac{q^2}{m}\mathbf{A}_\mathrm{ext}\cdot \mathbf{A}^{(2)}_{\mathrm{spin}}$ depict the interaction between the electronic charge $q$ and the spin dependent mean-field terms. They illustrate the mean-spin responses.

Let us focus first on the terms of type (A).

\begin{figure}[h!]
\begin{center}
\includegraphics[width=8.6cm]{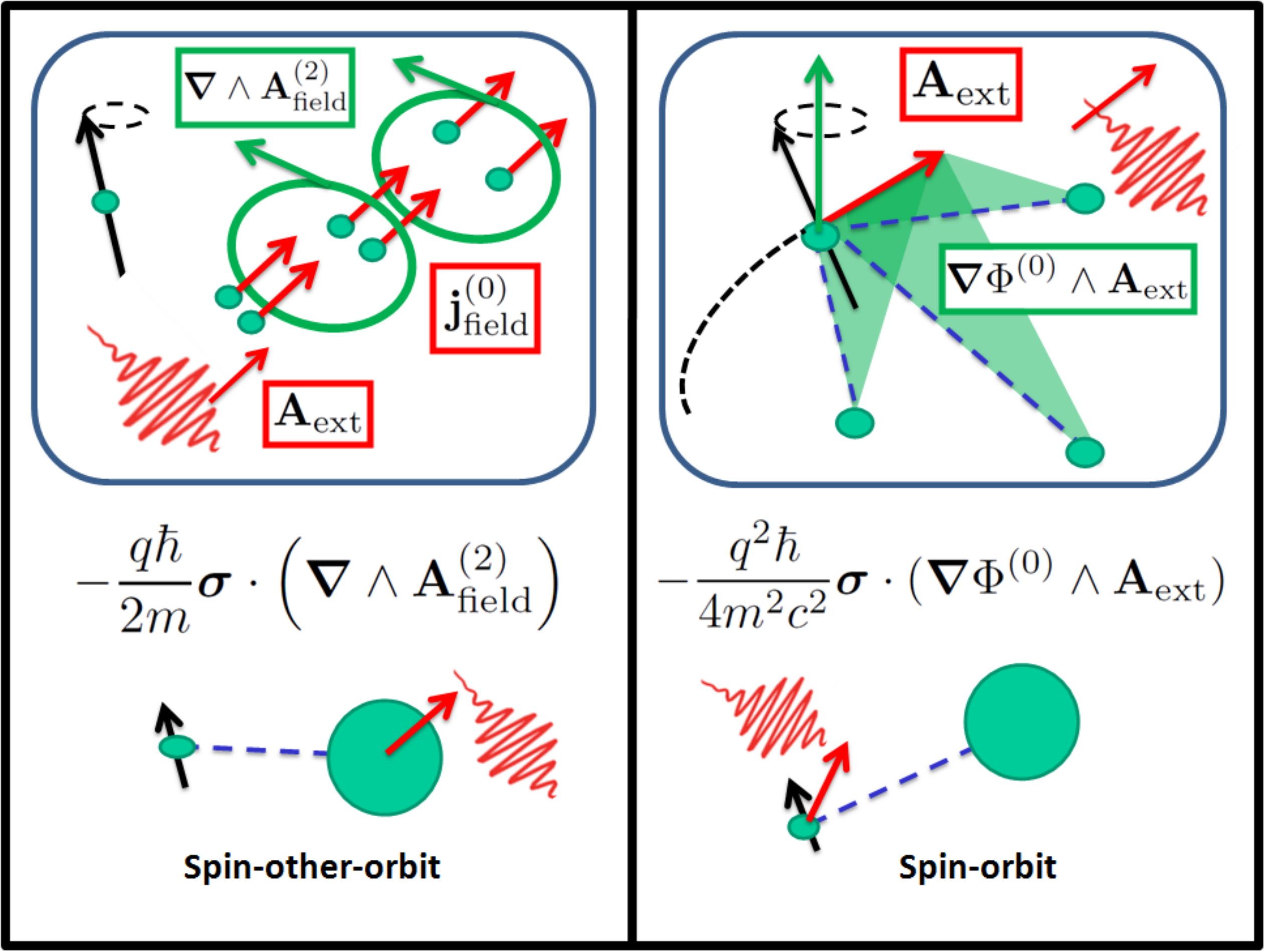}
\end{center}
\caption{(Color online) Mechanisms of type (A) (see details in text).}
\label{fig2}
\end{figure}

(A1) The Zeeman-like term $-\frac{q\hbar}{2m}\boldsymbol{\sigma }\cdot \left(\boldsymbol{\nabla}\wedge\mathbf{A}^{(2)}_{\mathrm{field}} \right)$ can be easily understood. The electron spin interacts with the magnetic field created by the motion of charges induced by the laser field. Indeed, the external vector potential $\mathbf{A}_\mathrm{ext}$ creates an internal current $\mathbf{j}_{\mathrm{field}}^{(0)}=-\frac{q}{m}\phi^{\dagger }\phi \mathbf{A}_{\mathrm{ext}}$ leading to a vector potential $\mathbf{A}^{(2)}_{\mathrm{field}}$ which is finally related to a light-induced magnetic field $\mathbf{B}_{\mathrm{eff}}=\boldsymbol{\nabla}\wedge\mathbf{A}^{(2)}_{\mathrm{field}} $. This intuitive picture is depicted on the left-panel of Fig. \ref{fig2}.

(A2) The spin-orbit mechanism $-\frac{q^2\hbar}{4m^{2}c^{2}}\boldsymbol{\sigma} \cdot ( \boldsymbol{\nabla }\Phi^{(0)} \wedge \mathbf{A}_\mathrm{ext})$ is the usual one involving the electron momentum $\mathbf{p}$ to which we add the momentum associated to the external vector potential $\mathbf{p}\mapsto \mathbf{p}-q\mathbf{A}_\mathrm{ext}$. The motion of the electronic charge is modified under the action of $q\mathbf{A}_\mathrm{ext}$, as well as its orbital angular momentum with respect to the positions of the other electrons. The effective magnetic field seen by the electron spin is therefore modified during the action of the pulse and reads $\boldsymbol{\nabla }\Phi^{(0)} \wedge \mathbf{A}_\mathrm{ext}$ (see right panel of Fig. \ref{fig2}).

\begin{figure}[h!]
\begin{center}
\includegraphics[width=8.6cm]{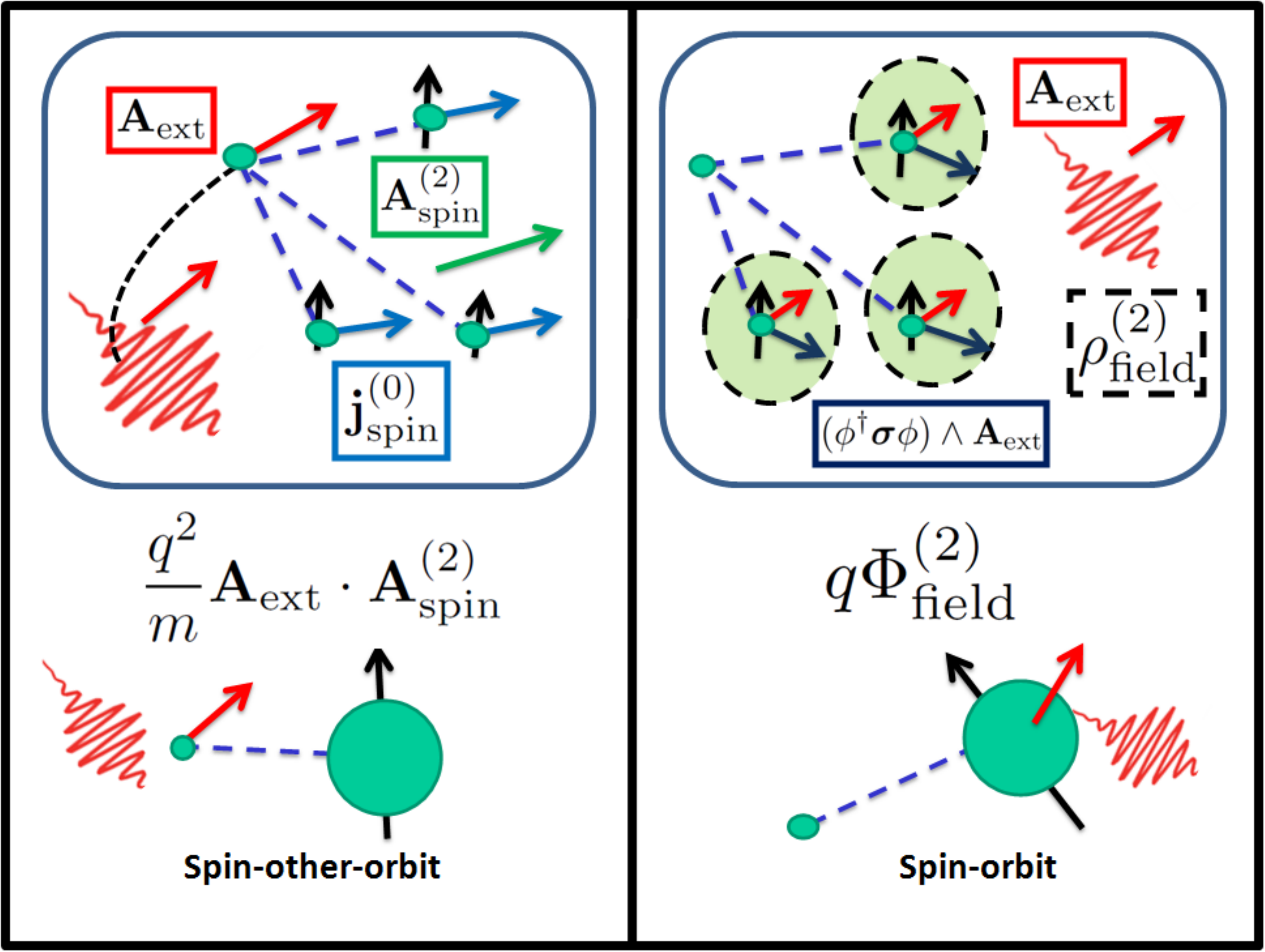}
\end{center}
\caption{(Color online) Mechanisms of type (B) (see details in text).}
\label{fig3}
\end{figure}

(B1) As for the second-type terms acting on the charge, one can see that the operator $\frac{q^2}{m}\mathbf{A}_\mathrm{ext}\cdot \mathbf{A}^{(2)}_{\mathrm{spin}}$ represents an electromagnetic energy involving two vector potentials: the one of the light $\mathbf{A}_\mathrm{ext}$ and one of the system $\mathbf{A}^{(2)}_{\mathrm{spin}}$. The latter is created by the internal spin current $\mathbf{j}_{\mathrm{spin}}^{(0)}$ as depicted in the left panel of Fig. \ref{fig3}. This interaction can be seen as an energy-term looking like $\frac{q^2 A^2}{2m}$ or as a paramagnetic dipolar coupling where $q\mathbf{A}_\mathrm{ext}$ substitutes to the electron momentum $\mathbf{p}$. We remember also that the magnetic field $\boldsymbol{\nabla } \wedge\mathbf{A}^{(2)}_{\mathrm{spin}} $ generated by $\mathbf{j}_{\mathrm{spin}}^{(0)}$ that couples to the electron spin via a Zeeman interaction corresponds to the internal spin-spin interaction given in $U^{\mathrm{int}}$.

(B2) Finally, the last term $q\Phi^{(2)}_{\mathrm{field}}$ is the most surprising. The electronic charge feels a second-order Hartree potential that is related to the charge density $\rho^{(2)}_{\mathrm{field}}=-\frac{q\hbar}{4m^2c^2}\boldsymbol{\nabla}\cdot\left((\phi^{\dagger }\boldsymbol{\sigma}\phi)\wedge \mathbf{A}_{\mathrm{ext}} \right)$, this latter quantity being a function of both the spin and laser field. The two vectors $\boldsymbol{\sigma}$ and $\mathbf{A}_{\mathrm{ext}}$ generate another one $\frac{}{}\left((\phi^{\dagger }\boldsymbol{\sigma}\phi)\wedge \mathbf{A}_{\mathrm{ext}} \right)$ which is finally associated to an effective electric field $\mathbf{E}_{\mathrm{eff}}$ (see right-panel of Fig. \ref{fig3}).

As a summary, the connection between these interactions and the microscopic sources producing the electromagnetic field is depicted in Table I. Given the order of magnitude of the coherent effects within current light-matter interaction conditions, and following the above analysis, one would suggest that these four mechanisms, may play an important role within the first few femtosecond of the demagnetization process.

\begin{table*}[ht!]
\caption{
Origin of the different types of interaction terms in the semi-relativistic light-induced mean field Hamiltonian $U^{\mathrm{int}}_{\mathrm{ext}}$ [Eq.(\ref{29})].}
\begin{ruledtabular}
\begin{tabular}{cccccc}
\hline
%sources $\setminus $ operators & Coulomb  & Darwin & spin-orbit& paramagnetic  & Zeeman \\
& \cr
 & Coulomb  & paramagnetic I & paramagnetic II& Zeeman  &  spin-orbit\\
 Sources
& $q\Phi^{(2)}$ & $-\frac{q}{m}\mathbf{A}^{(2)}\cdot\mathbf{p}$ & $\frac{q^2}{m}\mathbf{A}^{(2)}\cdot \mathbf{A}_\mathrm{ext}$ & $ - \frac{q\hbar}{2m}\boldsymbol{\sigma }\cdot (\boldsymbol{\nabla}\wedge\mathbf{A}^{(2)}) $  & $-\frac{q^2\hbar}{4m^{2}c^{2}}\boldsymbol{\sigma} \cdot \left(\boldsymbol{\nabla}\Phi^{(0)}\wedge \mathbf{A}_\mathrm{ext}\right)$\\
& & & &  &\\
\hline
$\rho^{(0)}$&  &    &  &  & spin-orbit [Eq. ({\ref{34}})] \\
 & &  &&  &\\
\hline
$\mathbf{j}^{(0)}_{\mathrm{orb}}$ &   & spin-free [Eq. ({\ref{32}})]& & & \\
& & & &  &\\
\hline
$\mathbf{j}^{(0)}_{\mathrm{spin}}$ &  &  & spin-other-orbit [Eq. ({\ref{36}})]&  &  \\
& & & &  &\\
\hline
$\mathbf{j}^{(0)}_{\mathrm{field}}$ &  & spin-free [Eq. ({\ref{30}})]& spin-free [Eq. ({\ref{31}})]& spin-other-orbit [Eq. ({\ref{33}})]& \\
& & & &  &\\
\hline
$\rho^{(2)}_{\mathrm{field}}$ & spin-orbit [Eq. ({\ref{35}})] &  &  &  &\\
& & & &  &\\
\hline
\end{tabular}
\end{ruledtabular}
\end{table*}
\section{Conclusions and perspectives}
In this work, the semi-relativistic limit of the self-consistent Dirac-Maxwell equations was obtained up to second-order in $1/c$ in the presence of an external electromagnetic field. The result consists on the availability of a self-consistent two-component mean field model that incorporates all the quantum and relativistic effects occurring at order $1/c^2$. The model also leads to a coherent light-induced semi-relativistic Hamiltonian that can describe the coherent interaction of an ultrafast laser pulse with a system of $N$ interacting electrons in the mean field approximation. The latter appears to be relevant to current laser-matter conditions. We have extracted four clearly identified mechanisms that involve the interaction of the external laser pulse with the spin degrees of freedom. They can be seen as light-induced spin-orbit interaction and light-induced spin-other-orbit interaction. We hope that the present work will lead to promising numerical investigations in a near future. Also, we believe that these results can be helpful to enlighten the issue of the microscopic interactions in the light-induced ultrafast spin dynamics. However, at this current step, the model presents several limitations that have to be incorporated in future analytical and numerical developments.

Firstly, one should be able to produce numerical calculations of the charge and spin dynamics and see how strong the light excitation should be to perturb significantly the equilibrium state fixed by the internal electromagnetic interactions \cite{Mor09,Mor14}. Furthermore, the set of obtained equations appears to be a time-dependent self-consistent Schrodinger-Poisson-like system including magnetic properties. It thus belongs to a well-know framework which can be in principle numerically solved \cite{Man10}. Another goal of such modeling is to establish a hierarchy between the mechanisms that involve the spin degrees of freedom, and to determine which mechanisms are the most relevant ones, depending of course on the initial conditions given by the choice of a physical system. Also, one should go beyond the Hartree approximation which neglects the exchange and correlations effects. The latter playing an important role in ferromagnetic materials, we hope to incorporate them in a future work. Finally, another improvement would be to incorporate a second (or multiple) light pulse(s) to describe efficiently the nonlinear optical effects. Indeed, most experimental techniques that extract information on the ultrafast spin and charge dynamics are based on nonlinear time-resolved pump-probe or four-wave mixing experiments, whose experimental signals exhibit a nonlinear combination of pump and probe beams intensities.

\section{Acknowledgments}
This work was supported in part by the Portuguese funding agency Fundac\~{a}o para a Ci\^{e}ncia e Tecnologia (FCT) through grant PTDC/FIS/122511/2010, co-funded by COMPETE and FEDER and by the project ATOMAG supported by the European Research Council advanced grant ERC-2009-AdG-20090325-247452.

\end{document}